\documentstyle[prb,aps,psfig,preprint]{revtex}
\newcommand \be {\begin{equation}}
\newcommand \bea {\begin{eqnarray}}
\newcommand \ee {\end{equation}}
\newcommand \eea {\end{eqnarray}}

\begin{document}

\title{Effective Charge and Spin Hamiltonian for the \\   
Quarter-Filled Ladder Compound $\alpha'$-NaV$_2$O$_5$} 

\author{Debanand  Sa\cite{e-dsa} and C. Gros}   
\address{Fachbereich Physik, University of the Saarland, 
66041 Saarbr\"ucken, Germany }

\date{\today}
\maketitle

\begin{abstract}
An effective intra- and inter-ladder charge-spin hamiltonian for the 
quarter-filled ladder 
compound $\alpha'$-NaV$_2$O$_5$ has been derived by using the standard 
canonical transformation method. In the derivation, it is clear that 
a finite inter-site Coulomb repulsion is needed to get a meaningful result 
otherwise the perturbation becomes ill-defined. Various limiting cases 
depending on the values of the model parameters have been analyzed in 
detail and the effective exchange couplings are estimated. We find  
that the effective intra-ladder exchange may become ferromagnetic for the 
case of zig-zag charge ordering in a purely electronic model.

We estimate the magnitude of the effective inter-rung Coulomb repulsion
in a ladder and find it to be about one-order of magnitude too small in 
order to stabilize charge-ordering. 

\bigskip 
\bigskip 
\bigskip 
\bigskip 
\bigskip 


\end{abstract} 
 

\section{Introduction}  
Low-dimensional quantum spin systems have received considerable attention 
from both theoretical as well as experimental point of view due to  
their unconventional physical properties. $\alpha'$-NaV$_2$O$_5$,    
which was believed to be a low-dimensional inorganic spin-Peierls (SP) 
compound \cite{iso} has recently been under intense investigation. 
$\alpha'$-NaV$_2$O$_5$ is an insulator and   
its magnetic susceptibility data fits very well to the one-dimensional 
Heisenberg chain model yielding an exchange interaction $J$=440 and 560 K 
for temperatures below and above the spin-Peierls transition temperature 
T$_{SP}$ (T$_{SP}$$\approx$ 34 K ) respectively \cite{iso,wei}.   
For T$\le$ T$_{SP}$, an 
isotropic drop in the susceptibility corresponding to a  
singlet-triplet gap of $\Delta_{SP }$=85 K has been observed.  

Recent X-ray structure data analysis \cite{Schnering98,smo}
at room temperature disfavours the  
previously reported non-centrosymmetric structure \cite{gal} 
($C_{2v}^7-P2_{1}mn$ space group) where V$^{+4}$ spin-1/2 ions form  
a one-dimensional Heisenberg chain, running along the crystallographic  
b-direction, separated by chains of V$^{+5}$ spin-zero ions.  
But the evidence for the centrosymmetric point group \cite{smo}   
($D_{2h}^{13}-Pmmn$) leads to only one type of V-site with a formal 
valence +4.5 in this compound. The V-sites then form a quarter-filled ladder,   
running along the b-axis with the rungs along the crystallographic a-axis.  
In the quarter-filled scenario, the electron spins are not localized at 
V-ions rather distributed over a V-O-V molecule which has found support 
by  NMR \cite{oha} as well as Raman measurements \cite{pop}.       

The nature of the state below T$_{SP}$ is presently
under intense investigation. Isobe and Ueda \cite{iso} 
originally proposed a usual 
spin-Peierls scenario but the detection of two inequivalent
V-sites in NMR \cite{oha} indicates a more complicated scenario
and the possibility of charge ordering. Several types of 
charge ordering, including 'in-line' \cite{tha} and
'zig-zag' \cite{seo,mos,gro} ordering has been proposed, but
only the zig-zag type of ordering has been found to be
in agreement with neutron scattering \cite{gro,Yoshimama98}
and anomalous X-ray scattering \cite{Nakao00}.

Recent determinations of the low-temperature crystal
structure found the space group Fmm2 \cite{lud,pal} 
and proposed the existence of 
three inequivalent V-ions below T$_{SP}$ 
\cite{lud,pal,Smaalen99}. This scenario was investigated
by DMRG (density-matrix renormalization group)
and a cluster-operator theory \cite{Gros00} and a strong
disagreement with neutron scattering data \cite{Yoshimama98}
was found. Ohama {\it et al.} recentely observed \cite{Ohama00}
that the aparent contradiction between cyrstallography
(three inequivalent V-ions bewlow T$_{SP}$) and NMR
(two inequivalent V-ions bewlow T$_{SP}$) could be resolved
when one considers possible subgroups of the originally proposed
space group Fmm2\cite{Ohama00}.

The crystal structure of $\alpha'$-NaV$_2$O$_5$ at T$>$T$_{SP}$ 
is orthorhombic 
(a=$11.318\,\mbox{\AA}$, b=$3.611\,\mbox{\AA}$, c=$4.797\,\mbox{\AA}$)
and consists of double chains of edge-sharing 
distorted tetragonal VO$_5$ pyramids running along the orthorhombic 
b-axis. These double chains are linked together via common corners 
of the pyramids and form layers.  
These are stacked along c-direction with 
no-direct V-O-V links. The Na atoms are located in between these 
layers. For the orbitals of the d-electrons at V-sites, those with d$_{xy}$ 
symmetry are suggested to be the relevant ones above and below T$_{SP}$ 
\cite{smo}. 
Due to the special orbital structure, the hopping amplitudes $t_a$ and $t_b$ 
are much larger than the inter-ladder hopping $t_{ab}$, $t_a$ and $t_b$ 
being the hopping amplitudes along the rung and the ladder direction 
respectively ($t_a\approx0.38\,\mbox{eV}$, 
$t_b\approx0.17\,\mbox{eV}$, $t_{ab}\approx0.012\,\mbox{eV}$). 
Since $\alpha'$-NaV$_2$O$_5$ is an insulator, it has been assumed 
that the on-site Coulomb repulsion $U$ is sufficiently large 
in comparision to the hopping amplitudes ($U\approx2.8\,\mbox{eV}$ 
from DFT calculation \cite{smo}). Moreover, 
one has to introduce the inter-site Coulomb repulsions, $V_a$, $V_b$, 
$V_{ab}$ to obtain the required charge ordering. In fact, it has been 
shown in a Hartree-Fock calculation \cite{seo} that the condition 
$U>V_a, V_b, V_{ab}> 
t_a, t_b, t_{ab}$ must be fulfilled in order to achieve a complete charge 
ordering. We consider this and other limits in the present paper. 

In the present work, we take into account the charge dynamics to obtain an 
effective low energy hamiltonian for $\alpha'$-NaV$_2$O$_5$. One starts 
from a pure electronic hamiltonian, which includes electron hopping in 
and between the ladders as well as the on-site and inter-site Coulomb 
interactions. The on-site Coulomb interaction $U$ is taken to be the largest 
parameter in our calculation. Since $\alpha'$-NaV$_2$O$_5$ is an insulator  
and we work at quarter-filling, one can 
project on a subspace of states which contains one electron per 
rung. Therefore, it is convenient to use an Ising pseudo-spin variable 
$\tau^z = \pm 1/2$ corresponding to a rung with an electron on
the {\it right/left} site of the rung. 
This is in the same spirit of Kugel and Khomskii's  
treatment of the orbital degeneracy problem in Jahn-Teller systems  
\cite{kug}. The spin and the 
pseudo-spin operators can be written as, 

\bea  
S^z={1\over 2}\sum_{\sigma} \sigma (R_{\sigma}^\dagger R_{\sigma}  
+L_{\sigma}^\dagger L_{\sigma}), \; 
S^+=R_{\uparrow}^\dagger R_{\downarrow}+L_{\uparrow}^\dagger L_{\downarrow}
,\;  
S^-=R_{\downarrow}^\dagger R_{\uparrow}+L_{\downarrow}^\dagger L_{\uparrow}    
, \; \label{S_op} \\   
\tau^z={1\over 2}\sum_{\sigma}(R_{\sigma}^\dagger R_{\sigma} 
-L_{\sigma}^\dagger L_{\sigma}), \; 
\tau^+=\sum_{\sigma}R_{\sigma}^\dagger L_{\sigma}, \;  
\tau^-=\sum_{\sigma}L_{\sigma}^\dagger R_{\sigma},   \label{tau_op}
\eea    

\noindent which for example yields, $R_{i,\uparrow}^\dagger L_{i,\uparrow} 
=\tau_i^+({1\over 2} +S_i^z)$,$\;$   
$R_{i,\uparrow}^\dagger L_{i,\downarrow}=\tau_i^+S_i^+$,$\;$    
$L_{i,\downarrow}^\dagger L_{i,\uparrow}=({1\over 2}-\tau_i^z)S_i^-$, etc.,   
where $\tau^{\pm}=\tau^x +i \tau^y$, $ S^{\pm} = S^x+i S^y$ and 
$R_{i,\sigma}^\dagger (L_{i,\sigma}^\dagger)$ are the creation operator 
of an electron with spin $\sigma$ on the right (left) site of the i-th 
rung of the ladder. In (\ref{S_op}) and (\ref{tau_op}) we have suppressed
the site-indices.


\section{Intra-ladder Exchange} 

\begin{figure}
\begin{center}

\psfig{file=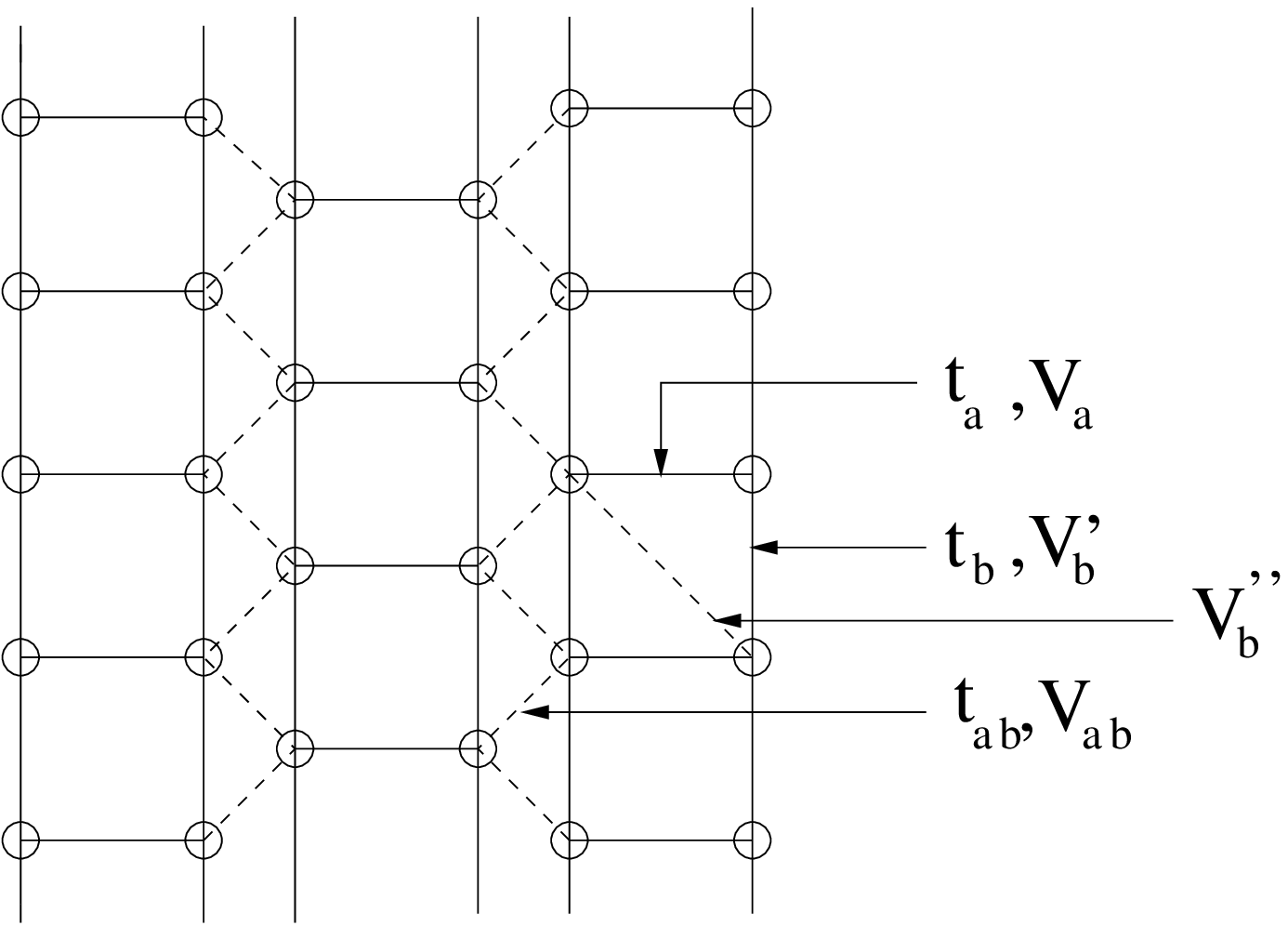}  
\medskip 
\caption{\label{Fig1}Schematic structure of $\alpha'$-NaV$_2$O$_5$ 
where the open 
circles stands for the Vanadium sites. Different parameters are 
also shown. }  

\end{center}
\end{figure}

Let us start with an electronic hamiltonian for the quarter-filled 
ladder (see Fig.\ \ref{Fig1}), which can be written as,  
$H = H_0 + H_0' + H_I$, with   

\be    
H_0= t_a \sum_{i,\sigma}(R_{i,\sigma}^\dagger L_{i,\sigma} + h. c.)  
+U \sum_{i} (n_{i,R\uparrow}  n_{i,R,\downarrow} + n_{i,L,\uparrow} 
n_{i,L,\downarrow}) 
+V_a \sum_{i,\sigma,\sigma'} 
n_{i,R,\sigma} n_{i,L,\sigma'},     
\ee  

\bea    
H_0'= V'_b \sum_{i,\sigma,\sigma'}
(n_{i,L,\sigma} n_{i+1,L,\sigma'}+n_{i,R,\sigma} n_{i+1,R,\sigma'} )  
+V''_b \sum_{i,\sigma,\sigma'}
(n_{i,L,\sigma} n_{i+1,R,\sigma'}+n_{i,R,\sigma} n_{i+1,L,\sigma'} )   
\nonumber \\  
+V_{ab} \sum_{<m,n>,\sigma,\sigma'} n_{m,R,\sigma} n_{n,L,\sigma'},   
\eea    

\be
H_I= t_b \sum_{i,\sigma}(R_{i,\sigma}^\dagger R_{i+1,\sigma} 
+L_{i,\sigma}^\dagger L_{i+1,\sigma} + h. c. )
+t_{ab} \sum_{<m,n>, \sigma}(R_{m,\sigma}^\dagger L_{n,\sigma}
+L_{n,\sigma}^\dagger R_{m,\sigma} ), \label{inter}     
\ee

\noindent where $t_a$, $U$ and $V_a$ are the hopping integral, on-site 
and the inter-site Coulomb repulsion in a rung respectively. 
$t_b$, $V'_b$ and $V''_b$ 
are the hopping integral and the Coulomb interaction in between rungs in a 
ladder, whereas $t_{ab}$ and $V_{ab}$ are the inter-ladder hopping and the  
Coulomb interaction respectively. 
$n_{i,R,\sigma}(n_{i,L,\sigma})$ is the electron density operator with  
spin $\sigma$ in the right (left) site of i-th rung and $<m,n>$ denotes 
the pair of rungs $m$ and $n$ on adjacent ladders.

We estimate the parameters of the inter-site Coulomb repulsion using a 
screened Coulomb repulsion, $V= e^2/(\epsilon d)$, 
where $\epsilon$ is the dielectric 
constant and $d$ the distance between the respective Vanadium atoms. 
The distance in between two V-ions along the rung and the leg of the
ladder (a- and b-directions) are $3.502\,\mbox{\AA}$ and 
$3.611\,\mbox{\AA}$. The dielectric constant is $\epsilon=11$ 
from microwave and far infrared measurements \cite{smi}. 
One obtains, $V_a=0.3738\,\mbox{eV}$ and $V'_b=0.3625\,\mbox{eV}$. 
The diagonal    
V-V distance in the b-direction in the ladder is $5.030\,\mbox{\AA}$, which    
implies, $V''_b=0.2643\,\mbox{eV}$. It will be clear from the 
later discussion   
that the effective inter-rung Coulomb repulsion $V_b$ is given by
the difference between $V'_b$ and $V''_b$, i.e.\ $V_b=V'_b-V''_b$, which
comes out to be small ($V_b=0.1023\,\mbox{eV}$) compared to $V_a$. 
Next, $V_{ab}=0.4305\,\mbox{eV}$, as
the V-V inter-ladder distance is $3.0401\,\mbox{\AA}$. Note that
$V_{ab}$ is slightly higher than $V_a$ and nearly four times higher 
than $V_b$.   
  
In order to develop a perturbation expansion, we start by
considering the case of a single two-leg quarter-filled 
ladder. 
The one-electron eigenstates of $H_0$ for a single rung
consist of bonding and anti-bonding wave functions, which we denote as 
$a^\dagger|0>={1\over \sqrt 2}(R^\dagger-L^\dagger)|0>$ and 
$s^\dagger|0>={1\over \sqrt 2}(R^\dagger+L^\dagger)|0>$,
with eigenenergies $-t_a$ and $+t_b$ respectively. 
Now let us consider the coupling of the rungs along the legs described 
by the first term in the hamiltonian $H_I$. 
In order to obtain a coupling between the pseudo-spin 
and the spin variables, we use here the standard canonical 
transformation method \cite{wol}, which is given by, 

\be  
H_{eff} = e^{iS} H e^{-iS},    
\ee 

\noindent where the operator $S$ is determined from the condition  

\be  
H_I + i [S,H_0] = 0,   
\ee  

\noindent which turns out to be  

\be  
\hat S = \sum_{n,n'} {i\over {(E_n'-E_n)}} |n><n|H_I|n'><n'| .         
\ee  

\noindent  Thus, the effective hamiltonian can be written as,  

\be  
H_{eff} = H_0 + H'_0   
- {1\over 2} \sum_{n,n',n''}\left({1\over {E_n'-E_n}} 
+{1\over {E_n'-E_n''}}\right) 
|n><n|H_I|n'><n'|H_I|n''><n''|,   
\ee  
 
\noindent where the initial and the final states $|n''>$ and $|n>$ 
are the two-rung states, 
i.e., all possible combinations of the bonding and the anti-bonding states 
between the nearest neighbour rungs. In the present case, there are sixteen 
possible such states which are the following:  

\be  
s_{i,\sigma}^\dagger s_{j,\sigma'}^\dagger |0>,\;    
s_{i,\sigma}^\dagger a_{j,\sigma'}^\dagger |0>,\;     
a_{i,\sigma}^\dagger s_{j,\sigma'}^\dagger |0>,\;    
a_{i,\sigma}^\dagger a_{j,\sigma'}^\dagger |0>, 
\ee   

\noindent with $\sigma, \sigma'=\uparrow, \downarrow$. 
The six intermediate states $|n'>$ which are the two-particle  
excited states in a rung, have to be antisymmetric   
under the exchange of both spin and pseudo-spin coordinates,  
in accordance with the Pauli principle. Thus, we have two sectors for the 
excited states depending on the total and the z-component of spin as well 
as the pseudo-spin quantum numbers which are labeled as, 
$|S,S^z;\tau,\tau^z>$. Hence, the states involved are, 
$|0,0;1,1>=R_{i,\uparrow}^\dagger R_{i,\downarrow}^\dagger |0>$,$\;$ 
$|0,0;1,-1>=L_{i,\uparrow}^\dagger L_{i,\downarrow}^\dagger |0>$,$\;$  
$|0,0;1,0>={1\over \sqrt 2}(R_{i,\uparrow}^\dagger L_{i,\downarrow}
^\dagger -R_{i,\downarrow}^\dagger L_{i,\uparrow}^\dagger)|0>$ $\;$ and  
$\;$ $|1,1;0,0>=R_{i,\uparrow}^\dagger L_{i,\uparrow}^\dagger |0>$, $\;$ 
$|1,-1;0,0>=R_{i,\downarrow}^\dagger L_{i,\downarrow}^\dagger |0>$,$\;$   
$|1,0;0,0>={1\over \sqrt 2}(R_{i,\uparrow}^\dagger L_{i,\downarrow}
^\dagger +R_{i,\downarrow}^\dagger L_{i,\uparrow}^\dagger)|0>$.  
The eigenenergies of the excited states in the large $U$ limit are 
$V_a$ for the spin-triplet states $|1,S^z;0,0>$
($S^z=-1,0+1$). For the 
spin-singlets, the eigenenergies are
$U'$ for ${1\over \sqrt 2}(|0,0;1,1>+|0,0;1,-1>)$ (symmetric),  
$U$ for ${1\over \sqrt 2}(|0,0;1,1>-|0,0;1,-1>)$ (antisymmetric) and 
$V'_a$ for $|0,0;1,0>$, with    
$U'\approx U+{{4 t_a^2}\over {U-V_a}}$ and 
$V'_a\approx V_a -{{4 t_a^2}\over {U-V_a}}$. 
After some lengthy but straightforward algebra and in the case 
of large but finite $U$, the total effective hamiltonian can be written as, 

\be
H_{eff}= H_0 + H'_0 + H_{eff}^{intra}+H_{eff}^{inter},    
\ee 

\noindent where $H_{eff}^{intra}$ is the effective intra-ladder hamiltonian  
which one can express as, $H_{eff}^{intra}=H_{eff}^{(t)}+ H_{eff}^{(s)}$, 
with $H_{eff}^{(t)}$ and $H_{eff}^{(s)}$ 
being the contribution due to the intermediate 
spin-triplet and spin-singlet states respectively.
$H_{eff}^{inter}$ is the effective inter-ladder hamiltonian which can 
be derived in a similar way and will be discussed in the next section. 
In terms of the pseudo-spin and spin 
variables, the unperturbed hamiltonian $H_0$ and $H'_0$ can be expressed as, 

\be 
H_0 = 2 t_a \sum_i \tau_i^x,   
\ee 

\be  
H'_0 = 2 (V'_b-V''_b)\sum_i \left({1\over4}+\tau_i^z \tau_{i+1}^z\right)
\,+\,N\,V_{b}''
+  V_{ab} \sum_{<m,n>} \left({1\over 4} - \tau_m^z \tau_n^z  
+{{\tau_m^z -\tau_n^z}\over 2}\right),
\label{H_0_prime}    
\ee  
where $N$ is the number of rungs.
In a similar way, the effective hamiltonian 
$H_{eff}^{(t)}$ can be written as,  

\be   
H_{eff}^{(t)}=-{{4 t_b^2}\over {V_a}} 
\sum_i \left({1\over 4} - \vec\tau_i\cdot\vec\tau_{i+1}\right)
\left({3\over4}+\vec S_i\cdot \vec S_{i+1}\right). \label{tf}    
\ee  

\noindent It is obvious from the 
above expression that $H_{eff}^{(t)}$ is independent of the Coulomb 
correlation energy $U$. This is due to the fact that while deriving 
this effective hamiltonian we have used the eigenenergies for the excited 
states which happen to be $V_a$ for this case.  
Since the effective hamiltonian $H_{eff}^{(s)}$ is obtained due to 
the contribution from the same intermediate states $|n'>$ and   
different initial and final states $|n>$ and $|n''>$, it can be  
written as,    
$H_{eff}^{(s)}= H_{eff}^{(s1)}+H_{eff}^{(s2)}+H_{eff}^{(s3)}$ (which 
are the contributions due to the antisymmetric, symmetric and 
$|0,0;1,0>$ intermediate states), with  

\be
H_{eff}^{(s1)}=- {{4 t_b^2}\over {U}}
\sum_i \left({1\over 4} -2\tau_i^x \tau_{i+1}^x 
+ \vec\tau_i\cdot\vec\tau_{i+1}\right)
\left({1\over 4}-\vec S_i\cdot \vec S_{i+1}\right), \label{s1}   
\ee

\bea
H_{eff}^{(s2)}=  -{{2 t_b^2}\over {U'-2t_a}}
\sum_i \left({1\over 4}+{{\tau_i^x+\tau_{i+1}^x}\over {2}} 
- 2\tau_i^y\tau_{i+1}^y 
+\vec \tau_i\cdot \vec \tau_{i+1}\right)
\left({1\over 4}-\vec S_i\cdot \vec S_{i+1}\right)   \nonumber \\
-{{2 t_b^2}\over {U'+2t_a}}
\sum_i \left({1\over 4} -{{\tau_i^x+\tau_{i+1}^x}\over {2}} 
-2\tau_i^y\tau_{i+1}^y 
+\vec \tau_i\cdot \vec \tau_{i+1}\right)
\left({1\over 4}-\vec S_i\cdot \vec S_{i+1}\right),  \label{s2}   
\eea

\bea
H_{eff}^{(s3)}=  -{{2 t_b^2}\over {V'_a -2t_a}}
\sum_i \left({1\over 4} +{{\tau_i^x+\tau_{i+1}^x}\over {2}} 
- 2\tau_i^z\tau_{i+1}^z
+\vec \tau_i\cdot \vec \tau_{i+1}\right)
\left({1\over 4}-\vec S_i\cdot \vec S_{i+1}\right)   \nonumber \\
-{{2 t_b^2}\over {V'_a +2t_a}}
\sum_i \left({1\over 4} -{{\tau_i^x+\tau_{i+1}^x}\over {2}} 
-2\tau_i^z\tau_{i+1}^z
+\vec \tau_i\cdot \vec \tau_{i+1}\right)
\left({1\over 4}-\vec S_i\cdot \vec S_{i+1}\right).  \label{s3}   
\eea

\noindent It should be noted here that one gets a non-zero contribution 
to $H_{eff}^{(t)}$ and $H_{eff}^{(s)}$ even if $U=\infty$ which will be 
discussed below. Moreover, it is obvious from the expression for 
$H_{eff}^{(t)}$ (see Eq.\ (\ref{tf})) that a finite $V_a$ is indeed 
needed to get a meaningful 
result otherwise the perturbation becomes ill-defined for $V_a=0$.


\subsection*{Limiting Cases and Discussion}  

{\bf Case I: $t_a =0$:} This limit implies $U'=U$ and $V'_a=V_a$ and thus,  
$H_{eff}^{(s1)}$ and $H_{eff}^{(s2)}$ can be combined together to yield, 

\be  
H_{eff}^{(s1)}+H_{eff}^{(s2)}=- {{8 t_b^2}\over {U}}  
\sum_i \left({1\over 4} +\tau_i^z \tau_{i+1}^z\right)
\left({1\over 4}-\vec S_i\cdot \vec S_{i+1}\right),  \label{s1s2}   
\ee  

\noindent whereas the effective hamiltonian $H_{eff}^{(s3)}$ gets 
reduced to,  

\be
H_{eff}^{(s3)}=-{{4 t_b^2}\over {V_a}} \sum_i
\left({1\over 4} + \vec\tau_i\cdot\vec\tau_{i+1}-2\tau_i^z\tau_{i+1}^z\right)
\left({1\over 4}-\vec S_i\cdot \vec S_{i+1}\right). \label{s3f}   
\ee

\noindent Since $H_{eff}^{(t)}$ neither depends on $t_a$ nor on $U$, it 
doesn't get affected in the above limiting case and the same is true for 
the other cases considered below. The effective hamiltonian  
derived by Thalmeier and Fulde \cite{tha} corresponds to
Eq.\ (\ref{s1s2}).   

{\bf Case II: $t_a=0$, $U=\infty$:} In this limit, which also corresponds
to the limit $V_a \gg 2t_a$, the contribution to 
the effective hamiltonian from $H_{eff}^{(s1)}$ and $H_{eff}^{(s2)}$ 
vanish and thus, the total intra-ladder effective hamiltonian 
becomes the sum of $H_{eff}^{(t)}$ and  $H_{eff}^{(s3)}$ 
(see Eq.\ (\ref{tf}) and (\ref{s3f})). This is   
what is exactly obtained by Mostovoy and Khomskii \cite{mos} but with 
a different interpretation.     
     
{\bf Case III: $U=\infty$, $V_a \ll 2t_a$:} In this case, the  
effective hamiltonian  $H_{eff}^{(s1)}$ and  $H_{eff}^{(s2)}$ vanish 
but $H_{eff}^{(s3)}$ reduces to, 

\be 
H_{eff}^{(s3)}= {{t_b^2}\over {t_a}} \sum_i
\left(\tau_i^x+\tau_{i+1}^x\right)
\left({1\over 4}-\vec S_i\cdot \vec S_{i+1}\right),  \label{s3f1}   
\ee 

\noindent so that the total effective intra-ladder hamiltonian 
becomes the sum of Eq.\ (\ref{tf}) and (\ref{s3f1}). Moreover, 
since $V_a \ll 2t_a$, the major contribution will be from Eq.\ (\ref{tf}). 

\begin{figure}
\begin{center}

\psfig{file=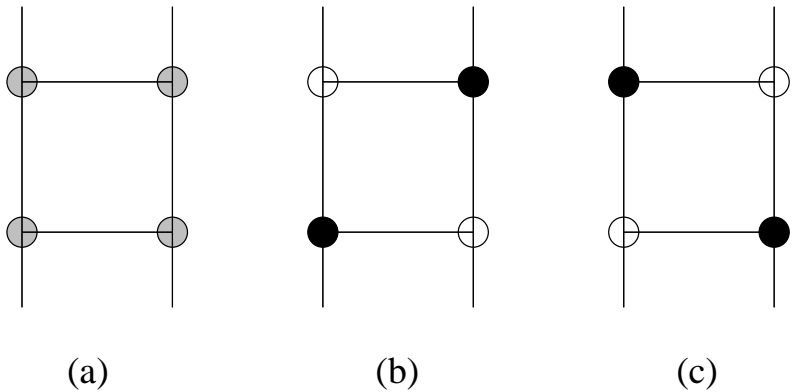}
\medskip 
\caption{\label{Fig2}Possible (a) disordered and (b), (c) completely 
charge ordered (zig-zag) configurations in a single ladder. 
The filled, shaded and open circles denote $V^{+4}$, $V^{+4.5}$
and $V^{+5}$ sites respectively.}

\end{center}
\end{figure}

{\bf Case IV:} {\it Disordered Phase:} In the disordered phase, where the 
electrons are in the bonding states (see Fig.\ \ref{Fig2} (a)), 
we can get  
an estimate of   
the effective exchange coupling in the effective hamiltonian by taking   
the averages over its charge (pseudo-spin) part. 
One can write down the effective 
exchange hamiltonian (disregarding the constant factors) as, 
$H_{eff}^{exch}= J_{eff}^{exch} \sum_i \vec S_i \cdot \vec S_{i+1}$. 
In the present case, we have,  $<\tau_i^x>=-1/2$ and  
$<\tau_i^y>=0=<\tau_i^z>$ whereas $<\tau_i^x\tau_{i+1}^x>=1/4$ and 
$<\tau_i^y\tau_{i+1}^y>=0=<\tau_i^z\tau_{i+1}^z>$. Thus, the effective 
exchange coupling due to the hamiltonian $H_{eff}^{(t)}$ and $H_{eff}^{(s1)}$ 
vanish, but that of $H_{eff}^{(s2)}$ and $H_{eff}^{(s3)}$ 
become, ${{-2t_b^2}\over{U'+2t_a}}$ and  ${{-2t_b^2}\over{V'_a+2t_a}}$ 
which yields $J_{eff}^{exch}={2t_b^2}({1\over{U'+2t_a}}+ 
{1\over{V'_a+2t_a}})$. It is clear that the exchange coupling here is 
antiferromagnetic ($>0$). Using the parameters mentioned in the 
present work, $J_{eff}^{exch}$ is estimated to be 0.08 eV. The expression for 
$J_{eff}^{exch}$ is exactly 
the same (for the case $V_a=0$) as obtained by Horsch and Mack \cite{hor}.  

{\bf Case V:} {\it Complete Charge Ordered Phase:}  We can get an estimate 
of the effective exchange couplings in the completely charge ordered  
(zig-zag) phase (see Fig.\ \ref{Fig2} (b, c)), following the same procedure as 
that of the disordered phase. Here, 
one has, $<\tau_i^x>=0=<\tau_i^y>$ and 
$<\tau_i^z>=1/2$,$\;$ $<\tau_{i+1}^z>=-1/2$ whereas $<\tau_i^x\tau_{i+1}^x>
=0=<\tau_i^y\tau_{i+1}^y>$ and  $<\tau_i^z\tau_{i+1}^z>=-1/4$. Hence, the 
effective exchange coupling due to $H_{eff}^{(s1)}$ and $H_{eff}^{(s2)}$ 
vanish but that of $H_{eff}^{(t)}$ and $H_{eff}^{(s3)}$ survive, which  
ultimately leads to $J_{eff}^{exch}={-2t_b^2}[{1\over{V_a}} 
-{1\over{2(V'_a-2t_a)}}-{1\over{2(V'_a+2t_a)}}]$. However, using the 
parameter values, it is calculated to be -0.17 eV. In the case where  
$U=\infty$ and for $0<(V_a /2t_a)<1$, $J_{eff}^{exch}$ becomes  
ferromagnetic ($<0$) whereas for $(V_a /2t_a)>1$, it is antiferromagnetic.   
The variation of $J_{eff}^{exch}$ with respect to the parameter 
$(V_a /2t_a)$ is shown in Fig.\ \ref{Fig4}. It is clear from the 
figure that there exists a minimum in the ferromagnetic region for 
$(V_a /2t_a)=1/ \sqrt 3$, where $J_{eff}^{exch,min}=-2.26\,({{t_b^2}/t_a})   
=-0.2\,\mbox{eV}$, which is quite large.


\section{Inter-ladder Exchange}  

Next, let us consider the hopping between the two nearest neighbour 
ladders, which is described by the second term of the hamiltonian $H_I$ 
(see Eq.\ (\ref{inter})), i.e.,   

\be
H_I^{inter}= t_{ab} \sum_{<m,n>,\sigma}(R_{m,\sigma}^\dagger L_{n,\sigma}
+L_{n,\sigma}^\dagger R_{m,\sigma}).  
\ee

\noindent The effective inter-ladder coupling between the charge and the 
spin degrees of freedom is derived in the same way as has been done for 
the single ladder case. Since the two-particle excited states in this 
case are exactly the same as what has been done earlier, the effective 
inter-ladder hamiltonian can be written as sum of two parts, i.e.,  
$H_{eff}^{inter} = H_{eff}^{inter(t)} + H_{eff}^{inter(s)}$, where the 
superscript `$t$' and `$s$' have the same meaning discussed in the 
previous section. The effective hamiltonian  
$H_{eff}^{inter(t)}$ is derived to be,     
 
\bea
H_{eff}^{inter(t)}=  -{{t_{ab}^2}\over {2(V_a -2t_a)}}
\sum_{<m,n>} \left({1\over 4}+{{\tau_m^x+\tau_n^x}\over {2}} 
-2\tau_m^y\tau_n^y  
+\vec \tau_m\cdot \vec \tau_n\right)\left({3\over 4} 
+\vec S_m\cdot \vec S_n\right) 
\nonumber \\
-{{t_{ab}^2}\over {2(V_a +2t_a)}}
\sum_{<m,n>} \left({1\over 4}-{{\tau_m^x+\tau_n^x}\over {2}} 
-2\tau_m^y\tau_n^y   
+\vec \tau_m\cdot \vec \tau_n\right)\left({3\over 4} 
+\vec S_m\cdot \vec S_n\right) 
\nonumber \\  
-{{t_{ab}^2}\over {V_a }}
\sum_{<m,n>} \left({1\over 4}-2\tau_m^x\tau_n^x
+\vec \tau_m\cdot \vec \tau_n\right)\left({3\over 4} 
+\vec S_m\cdot \vec S_n\right),    
\label{nt}  
\eea

\noindent whereas $H_{eff}^{inter(s)}$ can be written as 
$H_{eff}^{inter(s)}=H_{eff}^{inter(s1)}+H_{eff}^{inter(s2)}
+H_{eff}^{inter(s3)}$, with   

\bea 
H_{eff}^{inter(s1)}={-{t_{ab}^2}\over {2(U -2t_a)}}
\sum_{<m,n>} \nonumber \\       
\left({1\over 4}+{{\tau_m^x+\tau_n^x}\over {2}} 
+{{\tau_m^z-\tau_n^z}\over {2}}    
-2\tau_m^z\tau_n^z    
-\tau_m^x\tau_n^z+\tau_m^z\tau_n^x  
+\vec \tau_m\cdot \vec \tau_n\right)   
\left({1\over 4}-\vec S_m\cdot \vec S_n\right) \nonumber \\
-{{t_{ab}^2}\over {2(U +2t_a)}}
\sum_{<m,n>}     \nonumber \\   
\left({1\over 4}-{{\tau_m^x+\tau_n^x}\over {2}} 
+{{\tau_m^z-\tau_n^z}\over {2}}    
-2\tau_m^z\tau_n^z   
+\tau_m^x\tau_n^z-\tau_m^z\tau_n^x 
+\vec \tau_m\cdot \vec \tau_n\right)       
\left({1\over 4}-\vec S_m\cdot \vec S_n\right) \nonumber \\
-{{t_{ab}^2}\over {U}}
\sum_{<m,n>} \left({1\over 4}+{{\tau_m^z-\tau_n^z}\over {2}}
-\vec \tau_m\cdot \vec \tau_n\right)\left({1\over 4} 
-\vec S_m\cdot \vec S_n\right).      
\label{ns1}  
\eea  

\noindent The expression for $H_{eff}^{inter(s2)}$ is exactly the same  
as that of $H_{eff}^{inter(s1)}$ except that $U$ here is replaced by $U'$. 
On the other hand, $H_{eff}^{inter(s3)}$ is obtained as, 

\bea
H_{eff}^{inter(s3)}=  -{{t_{ab}^2}\over {2(V'_a -2t_a)}}
\sum_{<m,n>} \left({1\over 4}+{{\tau_m^x+\tau_n^x}\over {2}} 
-2\tau_m^y\tau_n^y
+\vec \tau_m\cdot \vec \tau_n\right)\left({1\over 4} 
-\vec S_m\cdot \vec S_n\right) 
\nonumber \\
-{{t_{ab}^2}\over {2(V'_a +2t_a)}}
\sum_{<m,n>} \left({1\over 4}-{{\tau_m^x+\tau_n^x}\over {2}} 
-2\tau_m^y\tau_n^y
+\vec \tau_m\cdot \vec \tau_n\right)\left({1\over 4} 
-\vec S_m\cdot \vec S_n\right) 
\nonumber \\
-{{t_{ab}^2}\over {V'_a }}
\sum_{<m,n>} \left({1\over 4}-2\tau_m^x\tau_n^x
+\vec \tau_m\cdot \vec \tau_n\right)\left({1\over 4} 
-\vec S_m\cdot \vec S_n\right).  
\label{ns3}  
\eea

\noindent Since the derivation of the effective inter-ladder hamiltonian 
proceeds in the same way as that of the intra-ladder case, the limiting 
cases will follow the same way as has been done before. Moreover, here  
also, one needs a finite $V_a$ (but $V_a\neq 2 t_a$) 
in deriving the effective hamiltonian, 
otherwise the perturbation becomes ill-defined for $V_a=0$. Furthermore, 
one gets a non-zero contribution to the effective inter-ladder hamiltonian 
even if $U=\infty$ .


\subsection*{Limiting Cases and Discussion}  
 
{\bf Case I: $t_a = 0$:} In this limit $U'=U$ and $V'_a=V_a$ follow  
naturally. Thus, $H_{eff}^{inter(t)}$ reduces to,    

\be  
H_{eff}^{inter(t)}=-{{2 t_{ab}^2}\over {V_a}}
\sum_{<m,n>} \left({1\over 4}+\tau_m^z\tau_n^z\right)   
\left({3\over 4}+\vec S_m\cdot \vec S_n\right).   \label{ntf}    
\ee  

\noindent  Similarly, $H_{eff}^{inter(s1)}$, $H_{eff}^{inter(s2)}$ and  
$H_{eff}^{inter(s3)}$ become,  

\be 
H_{eff}^{inter(s1)}+H_{eff}^{inter(s2)} =   
-{{4 t_{ab}^2}\over {U}}
\sum_{<m,n>} \left({1\over 4}+{{\tau_m^z-\tau_n^z}\over {2}}
-\tau_m^z\tau_n^z\right)  
\left({1\over 4}-\vec S_m\cdot \vec S_n\right), \label{ns1s2f}   
\ee  

\be 
H_{eff}^{inter(s3)}=-{{2 t_{ab}^2}\over {V_a}}
\sum_{<m,n>} \left({1\over 4}+\tau_m^z\tau_n^z\right)  
\left(1/4-\vec S_m\cdot \vec S_n\right). \label{ns3f}   
\ee   

\noindent The expression  Eq.\ (\ref{ns1s2f}) is the same one as 
has been obtained by Thalmeier and Fulde \cite{tha}. 

{\bf Case II: $t_a=0$, $U=\infty$:} In this case, the contributions  
from $H_{eff}^{inter(s1)}$and $H_{eff}^{inter(s2)}$ vanish.
The contributions from $H_{eff}^{inter(t)}$ and $H_{eff}^{inter(s3)}$ can be 
combined together to yield, 

\be  
H_{eff}^{inter}=H_{eff}^{inter(t)}+H_{eff}^{inter(s)}  
=-{{2 t_{ab}^2}\over {V_a}}
\sum_{<m,n>} \left({1\over 4}+\tau_m^z\tau_n^z\right),    
\ee

\noindent which becomes independent of the spin degrees of freedom. 

{\bf Case III: $U=\infty$, $V_a \ll 2 t_a $:} In this case,     
the contributions from $H_{eff}^{inter(s1)}$ and $H_{eff}^{inter(s2)}$ 
vanish.
Combining $H_{eff}^{inter(t)}$ and $H_{eff}^{inter(s3)}$, the spin-dependence
drops out and one obtains,  

\be 
H_{eff}^{inter}= H_{eff}^{inter(t)}+H_{eff}^{inter(s3)}= 
{{t_{ab}^2}\over {4 t_a }} \sum_{<m,n>} \left(\tau_m^x + \tau_n^x\right) 
-{{t_{ab}^2}\over {V_a }}\sum_{<m,n>} \left({1\over 4}-2\tau_m^x\tau_n^x 
+\vec \tau_m\cdot \vec \tau_n\right).    
\ee  

{\bf Case IV:} {\it Disordered Phase:} Following the procedure already 
mentioned for the intra-ladder case, we can get an estimate of the  
effective exchange coupling between the nearest neighbour ladders   
(see Fig.\ \ref{Fig3} (a)) by taking the averages over the charge part 
in the    
effective hamiltonian. Here also, one can write down the effective   
hamiltonian (disregarding the constant factors) as,
$H_{eff}^{inter,exch}= (J_{eff}^{inter,exch(t)}+J_{eff}^{inter,exch(s)}) 
\sum_{<m,n>}\vec S_m \cdot \vec S_n = J_{eff}^{inter,exch} \sum_{<m,n>}  
\vec S_m \cdot \vec S_n$. In the present case, one has,  
$<\tau_m^x>=<\tau_n^x>=-1/2$ and $<\tau_m^y>=<\tau_n^y>=0=
<\tau_m^z>=<\tau_n^z>$ whereas $<\tau_m^x\tau_n^x>=1/4$ and
$<\tau_m^y\tau_n^y>=0=<\tau_m^z\tau_n^z>$. Thus, the effective
exchange coupling which is due to $H_{eff}^{inter(t)}$,  
$H_{eff}^{inter(s1)}$, $H_{eff}^{inter(s2)}$ and $H_{eff}^{inter(s3)}$ is 
given as, $J_{eff}^{inter,exch}={{-t_{ab}^2}\over{2}} [{1\over {(V_a+2t_a)}}  
-{1\over{(U+2t_a)}}-{1\over{(U'+2t_a)}}  
-{1\over{(V'_a+2t_a)}}]$. The exchange coupling here is antiferromagnetic  
and is estimated to be $(J_{eff}^{inter,exch}/t_{ab}^2)=0.39$. 
For $V_a=0$, it gives rise to the same expression as obtained by   
Horsch and Mack \cite{hor}. 

{\bf Case V:} {\it Complete Charge Ordered Phase:} Here, we have 
four different completely charge ordered (zig-zag) phase depending on 
the state through which we compute the averages over the 
charge part of the effective hamiltonian. In all these cases, one has,
$<\tau_m^x>=<\tau_n^x>=0=<\tau_m^y>=<\tau_n^y>$ and
$<\tau_m^x\tau_n^x>=0=<\tau_m^y\tau_n^y>$. In addition to this, one has, 

\begin{description} 
\item[(i)] $<\tau_m^z>=1/2$,  $<\tau_n^z>=-1/2$ and
$<\tau_m^z\tau_n^z>=-1/4$, where the averages are due to the state 
$R_{m,\alpha}^\dagger L_{n,\beta}^\dagger |0>$ (Fig.\ \ref{Fig3} (b)). 
The effective exchange coupling for $H_{eff}^{inter(t)}$ and
$H_{eff}^{inter(s3)}$ vanish but that of $H_{eff}^{inter(s1)}$
and $H_{eff}^{inter(s2)}$ are finite which gives rise to,
$J_{eff}^{inter,exch}={t_{ab}^2} [{1\over {2(U-2t_a)}}+
{1\over {2(U+2t_a)}}+{1\over {U}}+{1\over {2(U'-2t_a)}} +
{1\over {2(U'+2t_a)}} + {1\over {U'}}]$. The exchange coupling here is
antiferromagnetic and is estimated to be $(J_{eff}^{inter,exch}/t_{ab}^2) 
=1.42$. For $t_a=0$, it gives rise to usual super-exchange, i.e.\ 
$J_{eff}^{inter,exch}={{4 t_{ab}^2}/U}$.   
\item[(ii)] $<\tau_m^z>=-1/2$,  $<\tau_n^z>=1/2$, and $<\tau_m^z\tau_n^z> 
=-1/4$, where the averages are taken over the state 
$L_{m,\alpha}^\dagger R_{n,\beta}^\dagger |0>$ (Fig.\ \ref{Fig3} (c)). 
In this case, $J_{eff}^{inter,exch(t)}=0=J_{eff}^{inter,exch(s)}$ which  
ultimately corresponds to the case $J_{eff}^{inter,exch}=0$.   
\item[(iii)] $<\tau_m^z>=<\tau_n^z>=1/2$ and $<\tau_m^z\tau_n^z>=1/4$,  
where the averages here are due to the state 
$R_{m,\alpha}^\dagger R_{n,\beta}^\dagger |0>$ (Fig.\ \ref{Fig3} (d)). 
The effective exchange couplings for $H_{eff}^{inter(s1)}$ and 
$H_{eff}^{inter(s2)}$ vanish. Thus, $J_{eff}^{inter,exch}$ is obtained from 
the contribution due to $H_{eff}^{inter(t)}$ and 
$H_{eff}^{inter(s3)}$ which is, 
$J_{eff}^{inter,exch}={-t_{ab}^2\over 2} [{1\over {2(V_a-2t_a)}} +
{1\over {2(V_a+2t_a)}}+ {1\over {V_a}} -
{1\over {2(V'_a-2t_a)}}-{1\over {2(V'_a+2t_a)}}-{1\over {V'_a}}]$.
The exchange coupling here is antiferromagnetic and is estimated to be
$(J_{eff}^{inter,exch}/t_{ab}^2)=2.8$. It vanishes for $U=\infty$ 
as well as for $t_a=0$.
\item[(iv)] $<\tau_m^z>=<\tau_n^z>=-1/2$ and $<\tau_m^z\tau_n^z>=1/4$,  
where the averages are taken over the state 
$L_{m,\alpha}^\dagger L_{n,\beta}^\dagger |0>$ (Fig.\ \ref{Fig3} (e)). 
Here, the effective exchange coupling turns out to be the same as that 
of (iii). 
\end{description}   

{\bf Case VI:} {\it A Phase with one Ladder Completely Charge Ordered 
and the Nearest Neighbour Disordered:} In this case, we have two 
possibilities, again depending on the state through   
which one computes the averages over the charge part in
the effective hamiltonian.  
In both the cases, one has, $<\tau_m^x>=0=<\tau_m^y>$,
$<\tau_n^y>=0=<\tau_n^z>$ and $<\tau_m^x\tau_n^x>=0=<\tau_m^y\tau_n^y>
=<\tau_m^z\tau_n^z>$.  Besides, one has, 

\begin{description} 
\item[(i)] $<\tau_m^z>=1/2$,
$<\tau_n^x>=-1/2$ and $<\tau_m^z\tau_n^x>=-1/4$, where the averages are 
taken over the state $R_{m,\alpha}^\dagger a_{n,\beta}^\dagger |0>$ 
(Fig.\ \ref{Fig3} (f)). All the effective
exchange couplings in this case turn out to be non-zero and hence
$J_{eff}^{inter,exch}$ is obtained as, $J_{eff}^{inter,exch}=
{-t_{ab}^2\over 2} {[{1\over {2(V_a+2t_a)}}+{1\over {2V_a}}
-{1\over {(U+2t_a)}}-{1\over {U}}-{1\over {(U'+2t_a)}}-{1\over {U'}}
-{1\over {2(V'_a+2t_a)}}-{1\over {2V'_a}}]}$.     
\item[(ii)] $<\tau_m^z>=-1/2$, $<\tau_n^x>=-1/2$ and $<\tau_m^z\tau_n^x>=1/4$ 
where the averages are due to the state 
$L_{m,\alpha}^\dagger a_{n,\beta}^\dagger |0>$ (Fig.\ \ref{Fig3} (g)). 
The effective exchange coupling which is due to $H_{eff}^{inter(t)}$ and
$H_{eff}^{inter(s3)}$ (the contributions due to $H_{eff}^{inter(s1)}$
and $H_{eff}^{inter(s2)}$ vanish) is given by, $J_{eff}^{inter,exch}= 
{-t_{ab}^2\over 2} {[{1\over {2(V_a+2t_a)}}+{1\over {2V_a}}
-{1\over {2(V'_a+2t_a)}}-{1\over {2V'_a}}]}$. 
\end{description}  

In both of the above mentioned 
cases, the exchange coupling becomes antiferromagnetic and is estimated 
to be $(J_{eff}^{inter,exch}/t_{ab}^2)=2.26$ for case (i) and 
1.65 for (ii). 
However, it vanishes in both cases for $U=\infty$ irrespective of 
whether $V_a \gg 2t_a$ or $V_a \ll 2t_a$.  
 

\section{Discussion and Conclusion}  

We have derived the effective spin-charge hamiltonian for
$\alpha'$-NaV$_2$O$_5$ for both intra-ladder and inter-ladder
exchange. We find a rich structure as a function of possible
realization of the microscopic parameters. We have found several,
in part unexpected, results.

\begin{figure}
\begin{center}

\psfig{file=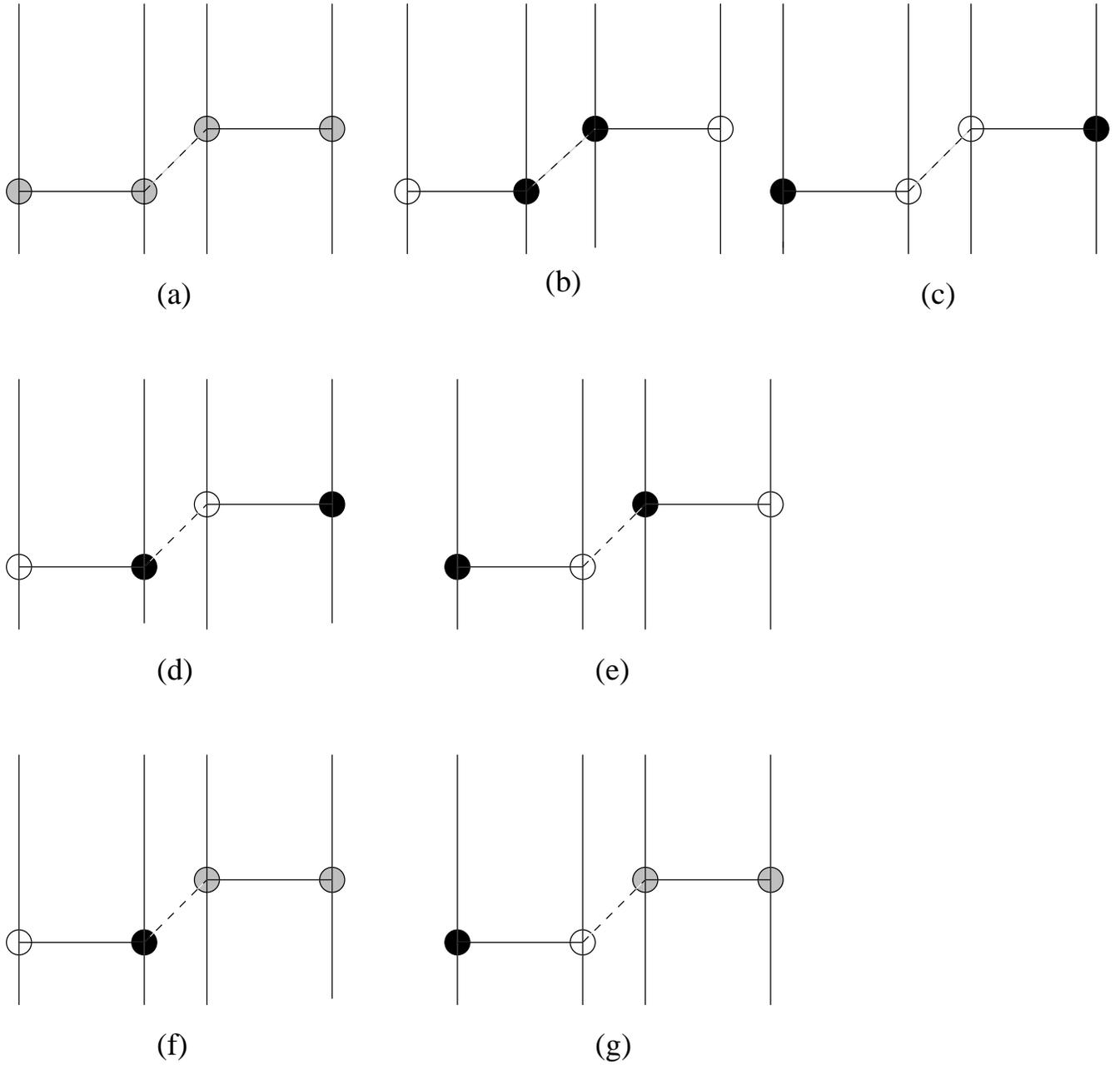}
\medskip 
\caption{\label{Fig3} Possible (a) disordered; (b), (c), (d), (e) 
completely charge ordered (zig-zag) and (f), (g) ordered-disordered  
configurations in between two adjacent ladders. }

\end{center}
\end{figure}

\begin{description}
\item[(i)] The effective magnetic exchange along the ladder
           decreases with increasing charge ordering (of
	   zig-zag type). For complete charge ordering, the
	   magnetic exchange becomes ferromagnetic and quite large
	   in magnitude.
\item[(ii)] The effective inter-ladder magnetic exchange in between 
            two given rungs of a charge-ordered
            and a charge-disordered ladder changes (only) by a
	    factor of two when the charge-density-wave is shifted by
	    a lattice constant along $a$ (compare 
	    Fig.\ \ref{Fig3} (f) and (g)).
\item[(iii)] There are novel terms of type $\tau_m^z \tau_n^x$ in the
            effective charge-charge inter-ladder interaction. These
	    terms, which are however rather small in magnitude, could
	    in principle stabilize a mixed charge-order configurations
	    like the one illustrated in Fig.\ \ref{Fig3} (f) and (g).
\end{description}

As a consequence of (i) and (ii), the proposed frustrated
spin-cluster model by de Boer {\it et al.} \cite{pal} seems to
be unlikely, since (a) no indications of ferromagnetic couplings
have been found experimentally and (b) the coupling in between
the one-dimensional spin-cluster chains of Ref.\ \cite{pal}
should be, as a consequence of (ii), rather strong. This conclusion
is consistent with a recent study of the frustrated spin-cluster
model by DMRG and a cluster-operator theory \cite{Gros00}.

\begin{figure}
\begin{center}

\psfig{file=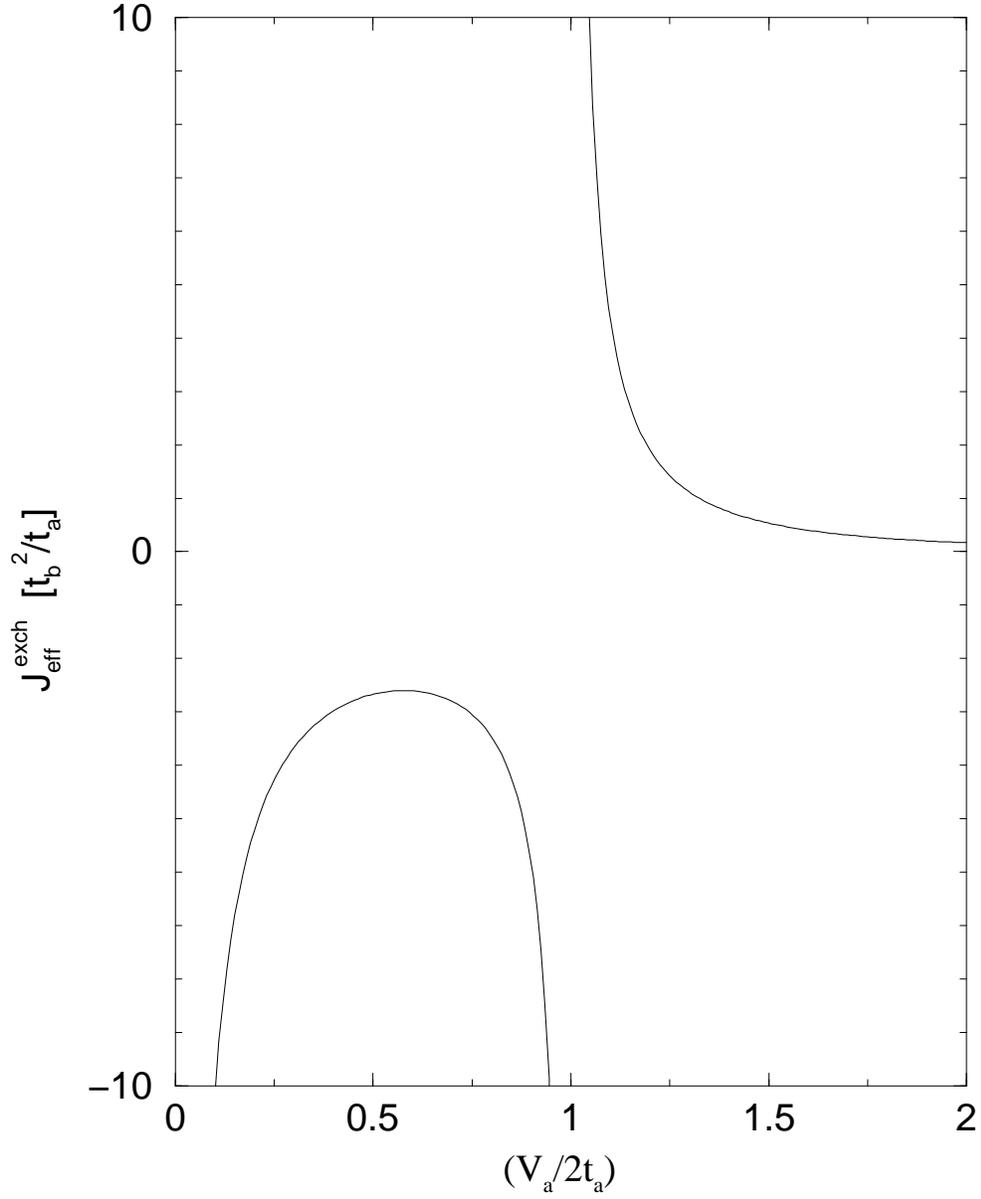,angle=-90}
\medskip 
\caption{\label{Fig4} Variations of the effective intra-ladder exchange 
coupling $J_{eff}^{exch}$ in units of $(t_b^2/t_a)$ with respect to 
the parameter $(V_a/2t_a)$ (for $U=\infty$). Note that the
perturbation expansion breaks down at $V_a=2t_a$.}  

\end{center}
\end{figure}

Let us note that the change in sign of the effective
intra-ladder magnetic exhange shown in Fig.\ \ref{Fig4}
cannot be described accurately by a perturbation expansion
in $t_b$. Near to the singularity at $V_a=2t_a$ the
perturbation expansion breaks down and the effective 
intra-ladder spin-hamiltonian becomes long-ranged.

Ultimately, the reason for the ferromagnetic 
intra-ladder coupling for the zig-zag ordering found in
the above calculation lies
in the fact that the charge-ordered state is not the
ground-state of $H_0$. In fact, the charge-ordered
state would be stabilized by $H_0'$ in the case of
a large effective inter-rung Coulomb-coupling $V_b'-V_b''$  
(see Eq.\ (\ref{H_0_prime})). For large values of
$V_b'-V_b''\gg 2t_a$, the perturbation expansion would 
yield (e.g.\ in mean-field approximation for $H_0'$)
an antiferromagnetic intra-ladder spin-spin coupling.
We did not show this calculation here, since our
estimated value for the inter-rung Coulomb repulsion
$V_b'-V_b''\approx0.1\,\mbox{eV}$ is about one-order
of magnitude too small in order to do the job. We therefore
believe that this small value of $V_b'-V_b''$ indicates (a)
the importance of elastic effects for the stabilization of
the observed phase transition at $T_c=34\,\mbox{K}$ and
(b) that the degree of charge ordering is far from 
complete. This is consistent with the proposal of
about 20\% charge ordering \cite{gro}.


\acknowledgements  The authors would like to thank R. Valent\'\i, 
J. V. Alvarez, F. Capraro and K. Pozgajcic for several discussions 
and critical reading of the manuscript.  
 


\end{document}